\documentclass[12pt,cites,graphicx]{article}
\parindent0mm\usepackage{graphicx}\title{Geometric Foundation of Thermo-Statistics,\\ 
Phase Transitions, Second Law of
Thermodynamics,\\ but without
Thermodynamic Limit\footnote{Presented at the 77$^{th}$ International Bunsen
Meeting ``Global Phase Diagrams'' Walberberg near K\"oln,
Germany, August 19. - 22. 2001}}
\author{D.H.E. Gross\\[3mm]
Berlin, Bereich Theoretische Physik,Glienickerstr.100\\ 14109
  Berlin, Germany and Freie Universit{\"a}t Berlin, Fachbereich
  Physik. \\E-mail: gross@hmi.de}
\begin{document}
\maketitle
\renewcommand{\thefootnote}{\fnsymbol{footnote}}
\newcommand{\cent}[1] {\begin{center}#1\end{center}}
\newcommand{\doublint}{\int\rule{-3.5mm}{0mm}\int} 
\newcommand{\mra}{\to} 
\newcommand{\vecbm}[1]{\mbox{\boldmath$#1$}}
\newcommand{\loo}{\,\raisebox{-.5ex}{$\stackrel{<}{\scriptstyle\sim}$}\,}
\newcommand{\goo}{\,\raisebox{-.5ex}{$\stackrel{>}{\scriptstyle\sim}$}\,}
\newcommand{\lra} {$\leftrightarrow$}
\newcommand{\vecb}[1]{\mbox{\bf#1}} 
\newcommand{\lora}{{\boldmath$\longrightarrow$}}
\noindent  A {\em geometric} foundation thermo-statistics is
presented with the {\em only axiomatic assumption of Boltzmann's
  principle} $S(E,N,V)=k\ln W$. This relates the entropy to the
geometric area $e^{S(E,N,V)/k}$ of the manifold of constant energy in
the ({\em finite}-$N$)-body phase space. From the principle, all
thermodynamics and especially {\em all phenomena of phase transitions
and critical phenomena can unambiguously be identified for even small
systems}. The topology of the curvature matrix $C(E,N)$ of $S(E,N)$
determines regions of pure phases, regions of phase separation, and
(multi-)critical points and lines. Phase transitions are linked to
convex (upwards bending) intruders of $S(E,N)$, where the canonical
ensemble defined by the Laplace transform to the intensive variables
becomes multi-modal, non-local, (it mixes widely different conserved
quantities). Here the one-to-one mapping of the Legendre transform
gets lost.  Within Boltzmann's principle, Statistical Mechanics
becomes a {\em geometric theory} addressing the whole ensemble or the
manifold of {\em all} points in phase space which are consistent with
the few macroscopic conserved control parameters. This interpretation
leads to a straight derivation of irreversibility and the Second Law
of Thermodynamics out of the time-reversible, microscopic, mechanical
dynamics. It is the whole ensemble that spreads irreversibly over the
accessible phase space not the single $N$-body trajectory.  This is
all possible without invoking the thermodynamic limit, extensivity,
or concavity of $S(E,N,V)$.  Without the thermodynamic limit or at
phase-transitions, the systems are usually not self-averaging, i.e.
do not have a single peaked distribution in phase space. The main
obstacle against the Second Law, the conservation of the phase-space
volume due to Liouville is overcome by realizing that a macroscopic
theory like Thermodynamics cannot distinguish a fractal distribution
in phase space from its closure.
\section{Introduction}
\label{intro}
There are many attempts to derive Statistical Mechanics from first
principles. The earliest are by
Boltzmann~\cite{boltzmann,boltzmann1877,boltzmann1884,boltzmann23a},
Gibbs~\cite{gibbs02,gibbs36}, and
Einstein~\cite{einstein02,einstein03,einstein04,einstein05a}. The two
central issues of Statistical Mechanics according to the deep and
illuminating article by Lebowitz~\cite{lebowitz99a} are to explain how
irreversibility (the Second Law of Thermodynamics) arises from fully
reversible, microscopic dynamics, and the other astonishing phenomenon
of Statistical Mechanics: the occurrence of phase transitions. 

In this paper I want to present an easy, straightforward derivation of
both aspects directly out of the microscopic time-reversal invariant
Newton-mechanics invoking a minimum of assumptions. We will see how
both problems are connected.There is an important aspect of
Statistical Mechanics which to my opinion was not sufficiently
considered up to now: Statistical Mechanics and also Thermodynamics
are {\em macroscopic} theories describing the {\em
  average}~\footnote{Here I do not speak of the {\em typical}
  behavior. This would only be the same if the system is {\em
    self-averaging}, which I do not demand, see below.}  behavior of
{\em all} $N$-body systems with the {\em same macroscopic
  constraints}.  It is this fact and nothing else that leads in a
simple and straightforward manner to the desired understanding of
irreversibility, the Second Law for {\em finite} $N$-body systems,
which obey a completely time reversible Hamiltonian dynamics, and
leads simultaneously to the full spectrum of phase-transition
phenomena.  It is certainly essential to deduce irreversibility from
reversible (here Newtonian) and not from dissipative dynamics as is
often done because just the {\em derivation of irreversibility} from
fully {\em reversible} dynamics is the main mystery of Statistical
Mechanics.  Here a first hint: Whereas a single trajectory in the
(finite-$N$)-body phase space returns after a {\em finite}
Poincar\'{e} recurrence time, a manifold of points develops in general
irreversibly with time, see below. This taken alone would not yet
allow for a rise of entropy. Since the entropy is the geometric
measure of the ensemble (see below) the Second Law seems to be in
conflict with Liouville's theorem which teaches us the invariance of
the phase-space volume. This contradiction is solved in section
\ref{fractSL} by defining a ``measure'' of the phase-space volume
which is more adequate to the redundant nature of a macroscopic theory
like Thermodynamics.\section{Minimum-bias deduction of Statistical
  Mechanics} Thermodynamics presents an economic but reduced
description of a $N$-body system with a typical size of $N\sim
10^{23}$ particles in terms of a very few ($M\sim 3-8$)
``macroscopic'' degrees of freedom ($dof$'s) as control parameters.
Here I will allow also for much smaller systems of some $100$
particles like nucleons in a nucleus. However, I assume that always
$6N\gg M$.  The belief that phase transitions and the Second Law can
exist only in the thermodynamic limit will turn out to be false.

Evidently, determining only $M$ $dof$'s leaves the overwhelming number
$6N-M$ $dof$'s undetermined. {\em All} N-body systems with the same
macroscopic constraints are {\em simultaneously} described by
Thermodynamics.  These systems define an {\em ensemble} $\cal{M}$ of
points~\footnote{In this paper I denote ensembles or manifold in phase
  space by calligraphic letters like ${\cal{M}}$.} in the $N$-body
phase space. Thermodynamics can only describe the {\em average}
behavior of this whole group of systems. I.e.\ it is a {\em
  statistical} or {\em probabilistic} theory. Considered on this level
we call Thermodynamics thermo-statistics or since Gibbs {\em
  Statistical Mechanics}. The dynamics of the (eventually interacting)
$N$-body system is ruled by its Hamiltonian $\hat{H}_N$. Let us in the
following assume that our system is trapped in an inert rectangular
box of volume $V$ and there is no further conservation law than the
total energy. The motion in time of all points of the ensemble follows
trajectories in $N$-body phase space $\{q_i(t),p_i(t)\}|_{i=1}^N$ (I
consider only classical mechanics) which will never leave the
($6N-1$)-dimensional shell (or manifold) $\cal{E}$ of constant energy
$E$ in phase space.  We call this manifold the {\em micro-canonical}
ensemble. An important information which contains the whole
equilibrium Statistical Mechanics including all phase transition
phenomena is the {\em area} $W(E,N) =:e^{S/k}$ of this manifold $\cal{E}$
in the $n$-body phase space.  Boltzmann has shown that $S(E,N,V)$ is
the {\em entropy} of our system.  Thus the entropy and with it
equilibrium thermodynamics has a {\em geometric} interpretation.

Einstein called Boltzmann's definition of entropy as e.g.\ written on
his famous epitaph
\begin{equation}
\fbox{\fbox{\vecbm{$S=k$\cdot$lnW$}}}\label{boltzmentr1}\end{equation}
{\em Boltzmann's principle}~\cite{einstein05d} from which Boltzmann was
able to deduce thermodynamics. Precisely $W$ is the number of
micro-states~\footnote{In the following I will call single points in
the $6N$-dim phase-space {\em states} or micro-states which are
specific microscopic realizations of the $N$-body system and
correspond to single N-body quantum states in quantum mechanics.
These must be distinguished from {\em macro-states} used in
phenomenological thermodynamics c.f.\ section~\ref{EPSformulation}.}
of the $N$-body system at given energy $E$ in the spatial volume $V$
and further-on I put Boltzmann's constant $k=1$:
\begin{eqnarray}
W(E,N,V)&=& tr[\epsilon_0\delta(E-\hat H_N)]\label{partitsum}\\
tr[\delta(E-\hat H_N)]&=&\int_{\{q\in V\}}{\frac{1}{N!}
\left(\frac{d^3q\;d^3p}
{(2\pi\hbar)^3}\right)^N\delta(E-\hat H_N)},\label{phasespintegr}
\end{eqnarray} 
$\epsilon_0$ is a suitable energy constant to make $W$ dimensionless,
the $N$ positions $q$ are restricted to the volume $V$, whereas the
momenta $p$ are unrestricted.  In what follows, I remain on the level
of classical mechanics. The only reminders of the underlying quantum
mechanics are the measure of the phase space in units of $2\pi\hbar$
and the factor $1/N!$ which respects the indistinguishability of the
particles (Gibbs paradox). With this definition,
eq.(\ref{boltzmentr1}), {\em the entropy $S(E,N,V)$ is an everywhere
  multiply differentiable, one-valued function of its arguments.} This
is certainly not the least important difference to the conventional
canonical definition. In contrast to
Boltzmann~\cite{boltzmann1877,boltzmann1884} who used the principle
only for dilute gases and to Schr\"odinger~\cite{schroedinger44}, who
thought equation (\ref{boltzmentr1}) is useless otherwise, I take the
principle as {\em the fundamental, generic definition of entropy}. In
a recent book~\cite{gross174} cf.\ also~\cite{gross173,gross175} I
demonstrated that this definition of thermo-statistics works well
especially also at higher densities and at phase transitions {\em
  without invoking the thermodynamic limit}. This is important: Elliot
Lieb~\cite{lieb97,lieb98a} considers the additivity of $S(E)$ and
Lebowitz~\cite{lebowitz99,lebowitz99a} the thermodynamic limit as
essential for the deduction of thermo-statistics. However, neither is
demanded if one starts from Boltzmann's principle. {\em Boltzmann's
  principle eq.(\ref{boltzmentr1}) is the only axiomatic assumption
  necessary for thermo-statistics.} {\em This is all that Statistical
  Mechanics demands}, no further assumption must be invoked. Neither
does one need extensivity \footnote{Dividing extensive systems into
  larger pieces, the total energy and entropy are equal to the sum of
  those of the pieces. I will call non-extensive systems where this is
  not the case in the following also ``Small'' systems~\cite{gross174}
  (with a capital $S$!) to stress the paradoxical point that the some
of largest systems in nature (globular star clusters) belong to this
group as well, nevertheless, they cannot be treated in the
thermodynamic limit.}, nor additivity, nor concavity of $S(E)$
c.f.~\cite{lavanda90}.

In the next section I will show how in contrast to the common claim of
textbooks Boltzmann's principle allows to define phase-transitions
unambiguously in ``Small'' non-extensive systems as well as in normal
``large'' extensive systems where our more general definition of phase
transitions (see below) will coincide with the conventional definition
by the Yang-Lee singularities~\cite{yang52,huang64}. Of course one
should not wonder if some familiar gospels of conventional canonical
thermo-statistics do not hold anymore in ``Small'' systems. This is
discussed in some more detail in subsection~\ref{equivalence}

``Small'' systems are either small many-body system like nuclei,
atomic clusters etc. where surface effects are important or the
largest systems possible like galaxies where the long-range gravity
does not allow to extrapolate to the thermodynamic limit.  Common to
all ``Small'' systems is that they are inhomogeneous. I.e.  the
fundamental {\em homogeneity} assumption of conventional
thermodynamics does not hold. Also at phase transitions of first order
do the systems become inhomogeneous. {\em Interfaces are the
  characteristic signal of the transition.}

Thermodynamics describes the development of {\em macroscopic} features
of many-body systems without specifying them microscopically in all
details. Therefore, traditional thermo-statistics works in the
thermodynamic limit of homogeneous infinitely large systems.  Why then
are we interested in ``Small'' systems? As will be explained later-on
small systems reveal deep peculiarities of statistical mechanics much
more sharply than macroscopic systems. Moreover, the overwhelming
majority of systems in nature are ``Small'' systems, e.g. all
astrophysical systems. The ``thermodynamic limit'' applies to some
$ccm$ but not to the really large ones.

\subsection{Why is the micro-canonical ensemble fundamental?} \label{why} 
During the dynamical evolution of a many-body system interacting by
short range forces the internal energy is conserved.  Only
perturbations by an external ``container'' can change the energy.
I.e.\ the fluctuations of the energy are
\begin{equation}
\frac{\Delta E}{E}\propto V^{-1/3},
\end{equation}
and for large volumes these energy fluctuations may be ignored.  If,
however, the diameter of the system is of the order of the range of
the force, i.e.\ the system is ``Small''or non-extensive, details of
the coupling to the container cannot be ignored.
\subsection{(Non)-equivalence of ensembles and self-averaging
\label{equivalence}} In contrast, the canonical ensemble does not 
care about these details, assumes the system is {\em homogeneous},
averages over a Boltzmann-Gibbs (exponential) distribution
$P_{BG}\{q_\alpha,p_\alpha\}=\frac{1}{Z(\beta)}e^{-\beta \hat
  H\{q_\alpha,p_\alpha\}}$ of energy and fixes only the mean value of
the energy by the temperature $1/\beta$. In order to agree with the
micro, $e^{-\beta E}W(E)$ must be sharp in $E$ i.e.  self-averaging,
which is usually not the case in non-extensive systems or at phase
transitions of first order.  Then one {\em must} work in the micro
ensemble.  The micro-ensemble assumes precise -- perhaps idealized --
boundary conditions for each particle independently of whether the
system is small or large. Therefore, already Gibbs considered the
micro-ensemble as the fundamental and the canonical as approximation
to it. He demonstrates clearly the failure of the canonical in cases
of phase separation or other situations where both ensembles differ,
footnote on page 75 of~\cite{gibbs02}, see
also~\cite{ehrenfest12,ehrenfest12a}.

There are important features where the micro-canonical statistics of
``Small'' systems deviates from the ``canonical'' structure of
conventional thermo-statistics of extensive systems in the
thermodynamical limit: E.g. {\em the familiar Legendre-transform
  structure, a paradigm of ``canonical'' thermo-statistics, is lost}.
Clearly, without self-averaging, fixing an intensive parameter like
the temperature $T$ does not fix the energy sharply.

Most evident example is a transition of first order in the canonical
ensemble at the transition temperature where the energy per particle
fluctuates by the specific latent heat even in the thermodynamic
limit. Related is the occurrence of {\em negative specific heat},
forbidden in the canonical thermodynamics, cf.
section~\ref{wrongcurv}, found in recent experiments on
nuclei~\cite{gross171,dAgostino00} which was predicted many years
before~\cite{gross95}. Here, there are at least three energies for the
same temperature c.f. section~\ref{systphas}. The present discussions
of non-extensive statistics as proposed by Tsallis~\cite{tsallis88} or
recently by Vives et al.~\cite{vives01} clearly miss this crucial
point. In the Tsallis statistics the entropy is expressed by the
mean-values of the extensive quantities like
$<\!\!E\!\!>$~\cite{vives01,martinez00a} controlled by a Lagrange
parameter $\beta$ or $\beta^*$, i.e. the Tsallis statistics works in
the canonical ensemble. Of course, this is equivalent to the
micro-ensemble only if the variance of the energy is small.  In one or
the other way the thermodynamic limit and self-averaging is still
demanded where Legendre transforms like $\beta\to E$ (may) become one
to one.  However, in the case of non-extensive systems the existence
of the thermodynamic limit is unlikely and so is the uniqueness of the
Legendre transformation.

\section{Phase transitions within Boltzmann's principle}
At phase-separation the system becomes {\em inhomogeneous} and splits
into different regions with different structure. {\em This is the main
  generic effect of phase transitions of first order.} Evidently,
phase transitions are foreign to the (grand-) canonical theory which
assumes homogeneous density distributions. Consequently, in the
conventional Yang-Lee theory, phase transitions~\cite{yang52} are
indicated by the zeros of the grand-canonical partition sum where the
grand-canonical formalism breaks down because of the Yang--Lee
singularities of the grand-canonical potentials like $[\ln Z(T,\mu)]$.
The micro-canonical formalism, esp. the micro-canonical entropy
$S(E,N,V)$ remains {\em continuous and multiple differentiable} at
phase transitions This is not the least important advantage of the
micro-formalism.

In the following I show in sharp contrast to a statement by
Schr\"odinger \cite{schroedinger44}, Boltzmann's principle to be
useful only for diluted gases, that Boltzmann's principle and the
micro-canonical ensemble gives a much more detailed and more natural
insight which moreover just corresponds to the experimental
identification of phase transitions by interfaces (inhomogeneities).

\subsection{Relation of the topology of {\boldmath$S(E,N,V)$} to the
Yang-Lee zeros of {\boldmath$Z(T,\mu,V$)}}

Yang-Lee singularities define phase transitions in the thermodynamic
limit. To explore the link, the grand-canonical partition sum may be
obtained out of the micro-canonical one by a double Laplace transform.
(In this limit it does not matter whether $N$ is discrete or
continuous.)\begin{eqnarray}
  Z(T,\mu,V)&=&\doublint_0^{\infty}{\frac{dE}{\epsilon_0}\;dN\;e^{-[E-\mu
      N-TS(E)]/T}}\nonumber\\
  &=&\frac{V^2}{\epsilon_0}\doublint_0^{\infty}{de\;dn\;e^{-V[e-\mu
      n-T
      s(e,n)]/T}}\label{grandsum}\label{laplace}\\
  &\approx&\frac{V^2}{\epsilon_0}\doublint_0^{\infty}de\;dn\;
  e^{-V[\mbox{\scriptsize const.+lin.+quadr.}]}\nonumber
\end{eqnarray}

The double Laplace integral (\ref{grandsum}) can be evaluated
asymptotically for large $V$ by expanding the exponent as indicated in
the third line to second order in $\Delta e,\Delta n$ around the
``stationary point'' $e_s,n_s$ where the linear terms vanish:
\begin{eqnarray}
\frac{1}{T}&=&\left.\frac{\partial s(e,n)}{\partial
e}\right|_{stat.point} \nonumber\\
\frac{\mu}{T}&=&-\left.\frac{\partial s(e,n)}{\partial n}\right|_{stat.point}
\label{statpoint}
\end{eqnarray} 
the only terms remaining to be integrated are the quadratic ones.

If the eigen-curvatures, $\lambda_1<0,\lambda_2<0,$ defined in
eqn.(\ref{curvdet}), and eqns.(\ref{statpoint}) have a single
solution ($e_s,n_s$), integral~(\ref{laplace}) is then a Gaussian
integral and yields:
\begin{eqnarray}
Z(T,\mu,V)&=&\frac{V^2}{\epsilon_0}e^{-V[e_s-\mu
n_s-T s(e_s,n_s)]/T}
\doublint_{-\infty}^{\infty}{dv_1\;dv_2\;
e^{V[\lambda_1 {\mbox{\boldmath$\scriptstyle
v$}}_1^2+\lambda_2{\mbox{\boldmath$\scriptstyle v$}}_2^2]/2}}\\
&=&e^{-(F(T,\mu,V)-\mu\bar{N})/T}.
\end{eqnarray}
We now investigate the specific free energy in the thermodynamic limit
$V\to\infty$:
\begin{equation}
 f(T,\mu,V)=
\frac{F(T,\mu,V)}{V}\to e_s-Ts_s
+\frac{T\ln{(\sqrt{-\lambda_1}\sqrt{-\lambda_2})}}{V}+
o(\frac{\ln{V}}{V})\label{asympt}.
\end{equation}
Here $\vecbm{v_1,v_2}$ are the eigenvectors in the \{$e,n$\}-plane
and $\lambda_1,\lambda_2$ the eigenvalues of the curvature matrix
with the determinant (Hessian):
\begin{equation} \det(e,n)= \left\|\begin{array}{cc}
      \frac{\partial^2 s}{\partial e^2}& \frac{\partial^2 s}
{\partial n\partial e}\\
      \frac{\partial^2 s}{\partial e\partial n}& \frac{\partial^2
        s}{\partial n^2}
\end{array}\right\|= \left\|\begin{array}{cc} s_{ee}&s_{en}\\
s_{ne}&s_{nn}
\end{array}\right\|=\lambda_1\lambda_2,\hspace{1cm}\lambda_1\ge\lambda_2
 \label{curvdet}
\end{equation}
$\lambda_1$ can be positive or negative. If $\lambda_1<0$ and
eqns.(\ref{statpoint}) have no other solution, the last two terms in
eqn.(\ref{asympt}) go to $0$ in the thermodynamic limit ($V\to\infty$),
and we obtain the familiar result for the free energy density:
\begin{equation}
f(T,\mu,V\to\infty)=e_s-Ts_s.
\end{equation} 
I.e.\ the {\em curvature $\lambda_1$ of the entropy surface $s(e,n,V)$
  or the largest eigenvalue of the curvature matrix decides whether
  the grand-canonical ensemble agrees with the fundamental
  micro-ensemble in the thermodynamic limit.} If this is the case and
eqns.(\ref{statpoint}) have a single solution or $s(e,n)$ touches its
concave hull at $e_s,n_s$, then there is a pointwise one to one
mapping of the micro-canonical entropy $s(e,n)$ to the grand-canonical
partition sum $Z(T,\mu)$, and $\ln[Z(T,\mu)]/V$ or $f(T,\mu)$ is
analytical in $z=e^{\beta\mu}$. Due to Yang and Lee we have then a
single, stable phase~\cite{huang64}. Otherwise, {\em the Yang-Lee
  zeros of $Z(T,\mu)$ reflect anomalous points/regions of
  $\lambda_1\ge 0$} \{$\det(e,n)\le 0$, in the cases studied here we
have always $\lambda_2<0$]\} where the {\em canonical partition sum does
  not reflect local properties of the micro-ensemble, i.e.\ does not
  respect the conservation laws, and mixes conserved quantities}.
This is crucial: As $\det(e_s,n_s)$ can be studied for finite or even
small systems as well, this is the only proper extension of phase
transitions to ``Small'' systems.

\subsection{The physical origin of the wrong
curvature\label{wrongcurv}} I will now discuss the physical origin of
the convex (upwards bending) intruders in the entropy surface for
systems with short-range coupling in two examples.

\subsubsection{Liquid-gas transition in sodium clusters \label{Nacluster}}
\begin{figure}[h]
\begin{minipage}[b]{9cm}
\includegraphics*[bb = 99 57 400 286, angle=-0, width=9cm,  
clip=true]{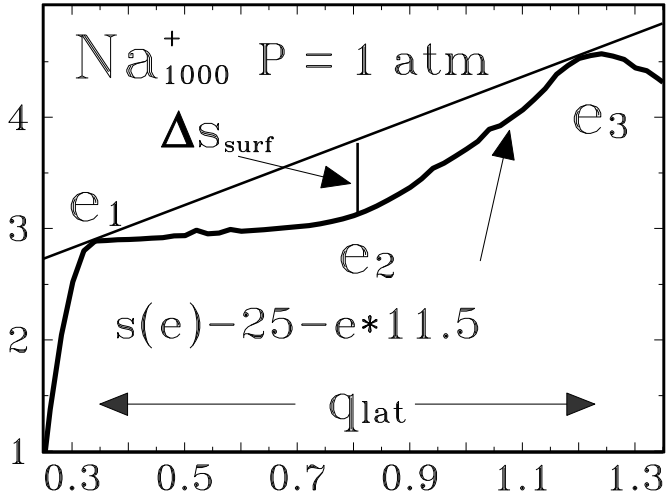}
\caption{MMMC~\protect\cite{gross174} simulation of the entropy 
  $s(e)$ per atom ($e$ in eV per atom) of a system of $N_0=1000$
  sodium atoms with an external pressure of 1 atm.  At the energy
  $e_1$ the system is in the pure liquid phase and at $e_3$ in the
  pure gas phase, of course with fluctuations. The
  latent heat per atom is $q_{lat}=e_3-e_1$. \label{naprl0f}}
\end{minipage}\hspace*{0.5cm}\begin{minipage}[b]{7.5cm}
  Attention: the curve $s(e)$ is artifically sheared by subtracting a
  linear function $25+e*11.5$ in order to make the convex intruder
  visible. {\em $s(e)$ is always a steeply monotonic rising function}.
  We clearly see the global concave (downwards bending) nature of
  $s(e)$ and its convex intruder. Its depth is the entropy loss due to
  the additional correlations by the interfaces. It scales $\propto
  N^{-1/3}$. From this one can calculate the surface tension per
  surface atom $\sigma_{surf}/T_{tr}=\Delta s_{surf}*N_0/N_{surf}$.
  The double tangent (Gibbs construction) is the concave hull of
  $s(e)$. Its derivative gives the Maxwell line in the caloric curve
  $T(e)$ at $T_{tr}$. In the thermodynamic limit the intruder would
  disappear and $s(e)$ would approach the double tangent from below.
  Nevertheless, the probability of configurations with
  phase-separations are suppressed by the (infinitesimal small) factor
  $e^{-N^{2/3}}$ relative to the pure phases and the distribution
    remains {\em strictly bimodal in the canonical ensemble}.
\end{minipage}
\end{figure}~\\
In the following table 
I compare the ``liquid--gas'' phase transition in sodium clusters of a
few hundred atoms with that of the bulk at 1 atm.\ c.f.\ also
fig.(\ref{naprl0f}). In these calculations \cite{gross157} we fixed
the sampling volume for each energy at that value where 
\begin{equation}
\left.\left (\frac{\partial S(E,V)}{\partial V}\right )\right/
\left (\frac{\partial S(E,V)}{\partial E}\right )= 1\mbox{ atm}.
\end{equation}
(This is in sharp contrast to Andersen's ``constant pressure
ensemble''~\cite{andersen80} where the volume can fluctuate at fixed
energy).  Conclusion: For systems with short range interactions a
convex intruder in $s(e,n)$ appears with the fragmentation of the
system into several clusters and monomers. The depth of the intruder
(surface entropy) scales with the number of surface particles. I.e.\ 
the convex intruder signals the preference of the system to become
inhomogeneous, which is the characteristic signal for the separation
of different phases (liquid and gas) at a phase transition of first
order.\\
\hspace*{0.5cm}
\begin{minipage}[h]{8cm}
\begin{tabular} {|c|c|c|c|c|c|} \hline 
&$N_0$&$200$&$1000$&$3000$&\vecb{bulk}\\ 
\hline 
\hline  
&$T_{tr} \;[K]$&$940$&$990$&$1095$&\vecb{1156}\\ \cline{2-6}
&$q_{lat} \;[eV]$&$0.82$&$0.91$&$0.94$&\vecb{0.923}\\ \cline{2-6}
{\bf Na}&$s_{boil}$&$10.1$&$10.7$&$9.9$&\vecb{9.267}\\ \cline{2-6}
&$\Delta s_{surf}$&$0.55$&$0.56$&$0.44$&\\ \cline{2-6}
&$N_{surf}$&$39.94$&$98.53$&$186.6$&$\vecbm{\infty}$\\ \cline{2-6}
&$\sigma/T_{tr}$&$2.75$&$5.68$&$7.07$&\vecb{7.41}\\
\hline
\end{tabular}
\end{minipage}\hspace*{1cm}\begin{minipage}[h]{7.5cm}~\\
  Table $1$ : ~Parameters of the liquid--gas transition of small sodium
  clusters (MMMC-calculation) in comparison with the bulk for rising
  number $N_0$ of atoms, $N_{surf}$ is the average number of surface
  atoms (estimated here as $\sum{N_{cluster}^{2/3}}$) of all clusters
  with $N_i\ge2$ together. $\sigma/T_{tr}=\Delta
  s_{surf}*N_0/N_{surf}$ corresponds to the surface tension. Its bulk
  value is adjusted to agree with the experimental values of the $a_s$
  parameter which we used in the liquid-drop formula for the binding
  energies of small clusters, c.f.  Brechignac et
  al.~\protect\cite{gross174}, and which are used in this calculation
  for the individual clusters.
\end{minipage}
~\\

\subsubsection{The global phase diagram portrayed by the topology of 
  the entropy surface {\boldmath$S(E,N)$}, here for Potts lattice gases} 

Having discussed in the previous example a system with a single
thermodynamic degree of freedom or control parameter (the energy $E$)
we will now study more subtle features.  If the system has two, or
more, control parameters, e.g. energy $E=Ve$ and particle number
$N=Vn$, where $V$ is the volume, we can have phase boundaries and
critical points. This reminds of the classical $P-V$ diagram of the
liquid--gas phase transition in Van-der-Waals theory. We are now also
able to identify multi-critical points. These were previously studied
in the canonical ensemble only, where sophisticated finite size
scaling is needed to identify these points. As example we investigate
the 3-states diluted Potts model on a {\em finite} 2-dim (here
$L^2=50^2$) lattice with periodic boundaries (to minimize effects of
the external surfaces of the system). The model is defined by the
Hamiltonian:
\begin{eqnarray}
H&=&-\sum_{i,j \in n.n.pairs}o_i o_j\delta_{\sigma_i,\sigma_j}\\
n&=&L^{-2}N=L^{-2}\sum_io_i .\nonumber
\end{eqnarray}

\begin{figure}[h]
\includegraphics*[bb =0 0 290 180, angle=0, width=14.5cm, 
clip=true]{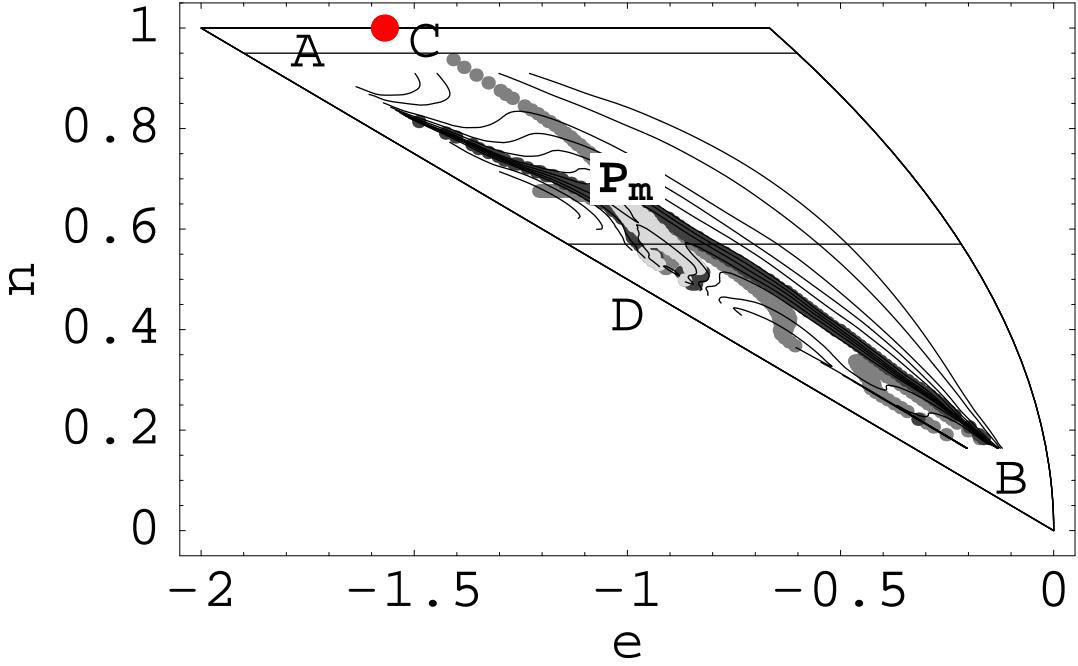}
\caption{Global phase diagram or conture plot of the curvature 
  determinant, eqn.(\ref{curvdet}), of the 2-dim Potts-3 lattice gas
  with $50*50$ lattice points, $n$ is the number of particles per
  lattice point, $e$ is the total energy per lattice point; Dark grey
  lines: $\det=0$, boundary of the region of phase coexistence
  ($\det<0$) in the triangle $AP_mB$; Light grey lines: minimum of
  $\det(e,n)$ in the direction of the largest curvature
  ($\vecbm{v}_{\lambda_{max}}\cdot\vecbm{\nabla}\det=0$), lines of
  second order transition; In the triangle $AP_mC$ pure ordered
  (solid) phase ($\det>0$); Above and right of the line $CP_mB$ pure
  disordered (gas) phase ($\det>0$); The crossing $P_m$ of the
  boundary lines is a multi-critical point. It is also the critical
  end-point of the region of phase separation ($\det<0$). The light
  gray region around the multi-critical point $P_m$ corresponds to a
  flat (cylindric) region of $\det(e,n)\sim 0$ and
{\boldmath$\vecbm{\nabla}$} \mbox{$\lambda_1$}{\boldmath$\sim 0$},
details see \protect\cite{gross173}; $C$ is the analytically known
position of the critical point which the ordinary $q=3$ Potts model
(without vacancies) would have {\em in the thermodynamic limit}
$N\to\infty$.} \label{det}
\end{figure}
Each lattice site $i$ is either occupied by a particle with spin
$\sigma_i =-1,0,\mbox{ or }1$, or it is empty (vacancy).  The sum is
over pairs of neighboring lattice sites $i,j$, and the occupation
numbers are:
\begin{equation}
o_i=\left\{\begin{array}{cl}
1&\mbox{, spin particle in site }i\\
0&\mbox{, vacancy in site }i\\
\end{array}\right. .
\end{equation}

This model is an extension of the ordinary ($q=3$)-Potts model to
allow also for vacancies. 

How to understand the line $CP_m$ of second-order transition? At zero
concentration of vacancies ($n=1$), we know that the system has in the
thermodynamic limit ($N\to\infty$) a continuous phase transition at
$e_c=1+\frac{1}{\sqrt{q}}\approx 1.58$ \cite{baxter73,pathria72}. With
rising number of vacancies the probability decreases to find a pair of
particles at neighboring sites with the same spin orientation. I.e.
this is analog to a larger number $q_{eff}$ of spin orientation on
each lattice site in the ordinary (completely filled) Potts model. We
know that there the transition of second order becomes a transition of
first order for $q>4$. Similarly, the inclusion of vacancies has the
effect of an increasing effective $q_{eff}\ge 3$.  This results in an
increase of the critical energy of the continuous phase transition
with decreasing $n$ and provides a line of continuous transition,
which is supposed to terminate when $q_{eff}$ becomes larger than $4$.
From here on the transition becomes first order. At smaller energies
the system is in one of the three ordered phases (spins predominantly
parallel in one of the three possible directions). Figure (\ref{det})
shows clearly how for a small system of $50*50$ lattice points all
phenomena of phase transitions can be studied from the topology of the
determinant of curvatures (Hessian~\ref{curvdet}) in the
micro-canonical ensemble. [In this example the second curvature
is always $\lambda_2<0$ and consequently,
$\mbox{sign}(\det)=-\mbox{sign}(\lambda_1)$]. 
\subsection{Conclusion and systematics of phase transitions in the
  micro-ensemble~\label{systphas}}

Now we can give a systematic and generic classification of phase
transitions in terms of the topology of curvatures of $s(e,n)$ which
applies also to ``Small'' systems:
\begin{itemize}
\item \begin{minipage}[t]{10cm} A {\bf single stable} phase by
    $\lambda_1<0$. Here $s(e,n)$ is locally concave (downwards bended)
    in both directions and eqns.(\ref{statpoint}) have a single
    solution $e_s,n_s$. Then there is a one to one mapping of the
    grand-canonical \lra the micro-ensemble. The order parameter is
    the direction $\vecbm{v}_1$ of the eigenvector of largest
    curvature $\lambda_1$. In many situations one may have only
    locally $\lambda_1<0$, but there may be further solutions to
    eqns.(\ref{statpoint}) farther away. Such cases have no equivalent
    in the canonical ensemble, here we will still speak of regions in
    \{$e,n$\} of pure phases embedded in regions of phase-separation.
\end{minipage}~~~\begin{minipage}[t]{7cm} 
\includegraphics*[bb = 61 6 429 624, angle=-90, width=5.5cm,
  clip=true]{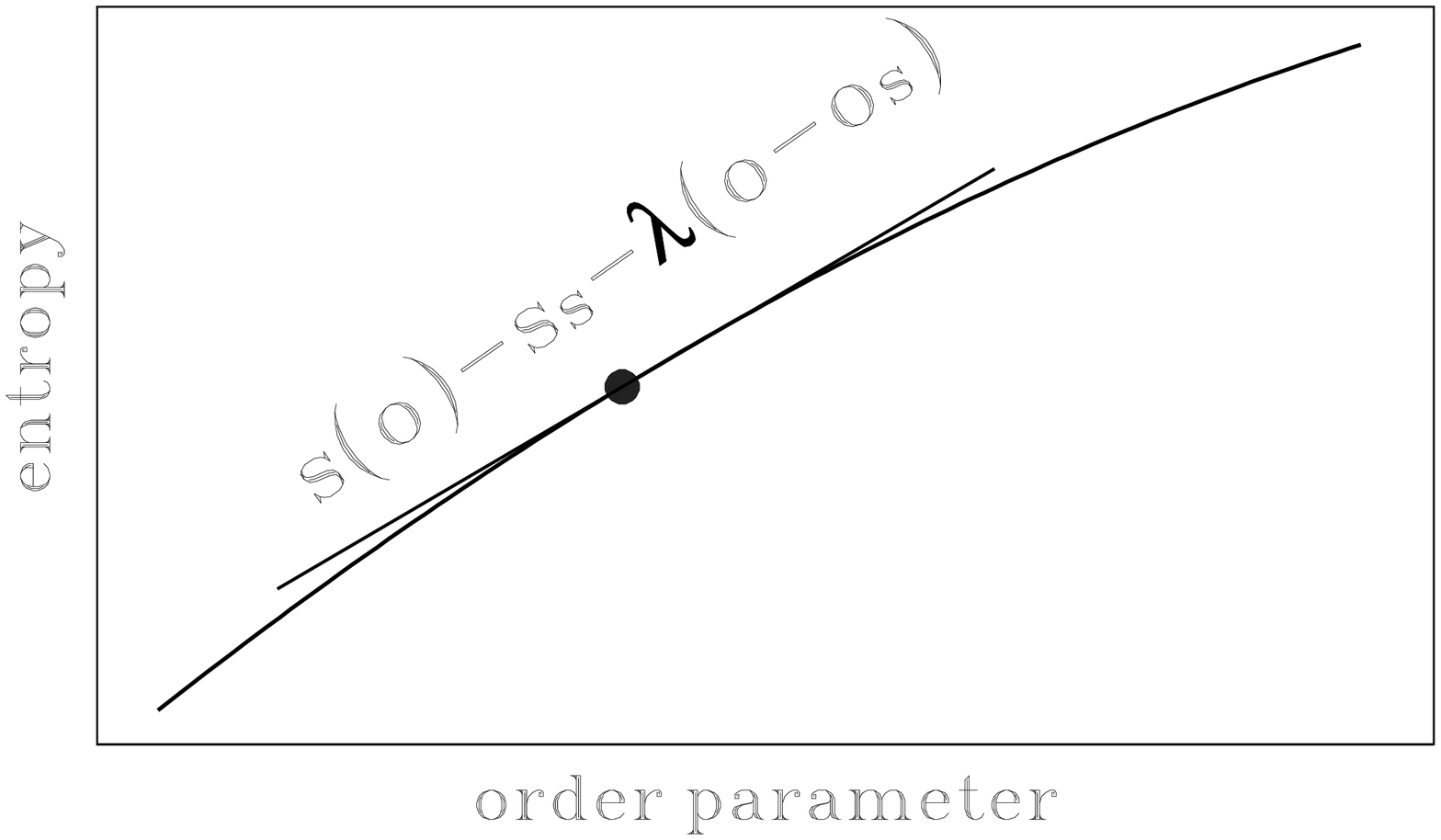}\end{minipage}~\\
\item \begin{minipage}[t]{10cm} A {\bf transition of first order} with
    phase separation and surface tension (c.f.subsection
    \ref{Nacluster}) indicated by $\lambda_1>0$. $s(e,n)$ has a convex
    intruder (upwards bended) in the direction $\vecbm{v}_1$ of the
    largest curvature. Then eqns.(\ref{statpoint}) have multiple
    solutions, at least three. The system is in the pure liquid phase
    at $o_1$ and in the pure gas phase at $o_3$. The whole convex area
    of \{e,n\} is mapped into a single point ($T,\mu$) in the
    grand-canonical ensemble (non-locality)\label{convex}. I.e.\ if
    the largest curvature of $S(E,N)$ is $\lambda_1>0$ {\em both
      ensembles are not equivalent, the (grand-) canonical ensemble is
      non-local in the order parameter and violates basic conservation
      laws.} C.f.~\cite{gross174,gross173,gross175,gross176}.
     \end{minipage}~~~\begin{minipage}[t]{7cm} 
       \includegraphics*[bb = 91 0 472 651, angle=-90, width=6cm,
       clip=true]{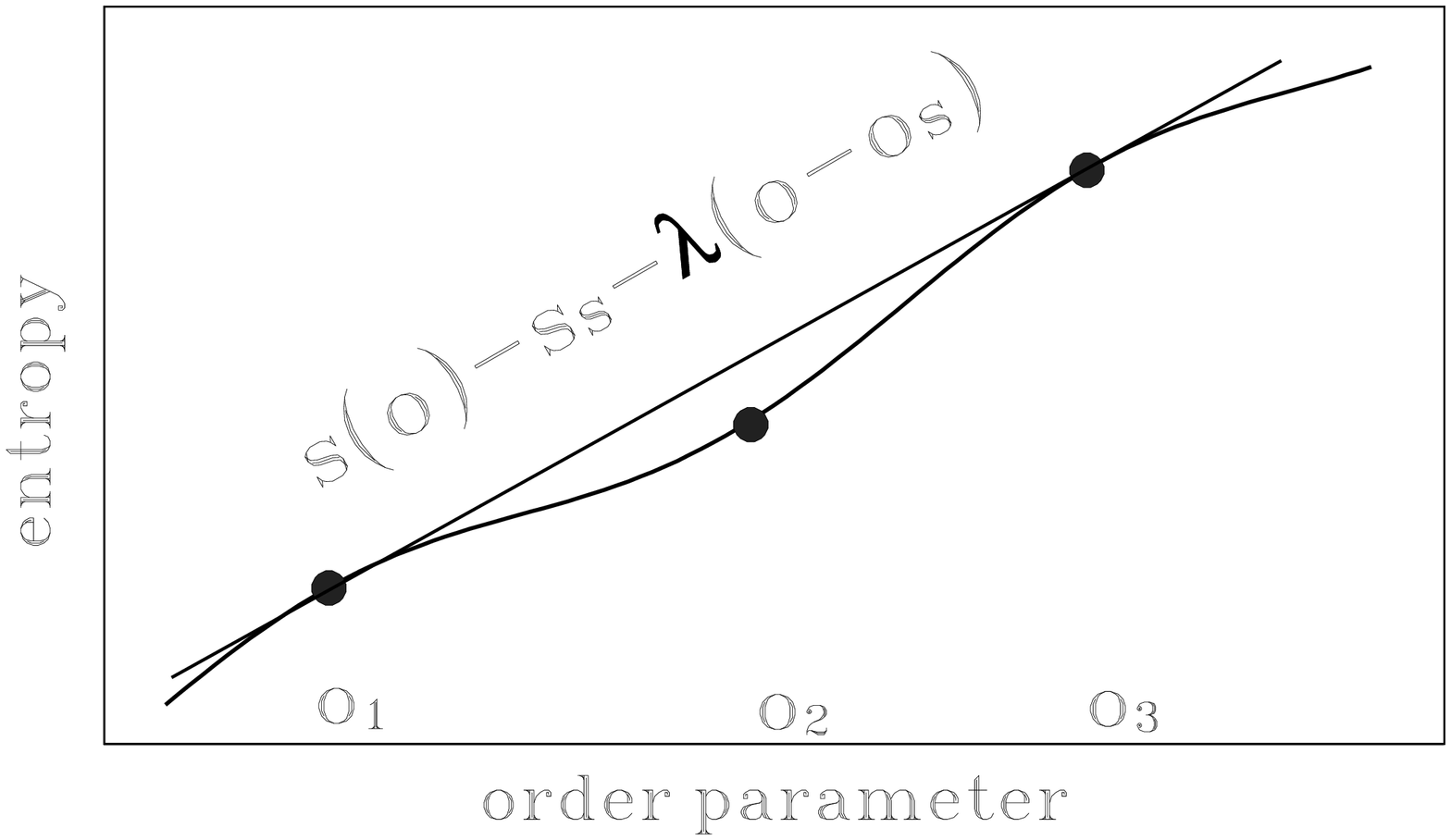} \end{minipage}\\
     The region in the plane of conserved control-parameters $e,n$
     where we have a separation of different phases, where
     $\lambda_1(e,n)>0$, is bounded by lines with $\lambda_1(e,n)=0$.
     Here one of the two coexisting phases gets depleted.  A special
     point on this boundary is the end-point of the transition of
     first order,
   \item where we have a {\bf continuous (``second order'')}
     transition with vanishing surface tension, where two neighboring
     phases become indistinguishable. This is at points where the two
     stationary points $o_1,o_3$ move into one another to become the
     critical end-point of the first order transition. This is then
     also a maximum of $\lambda_1$. I.e.\ where $\lambda_1(e,n)=0$ and
     $\vecbm{v}_{\lambda_1=0}\cdot\vecbm{\nabla}\lambda_1=0$.  These
     are the {\em catastrophes} of the Laplace transform $E\to T$.
     Here $\vecbm{v}_{\lambda_1=0}$ is the eigenvector of the
     curvature matrix belonging to the largest curvature eigenvalue
     $\lambda_1=0$.  ($\vecbm{v}_1$ plays the role of the order
     parameter of the transition. In this direction one moves fastest
     from one phase to the other.)  Furthermore, there may be also
     whole lines of second-order transitions like the line $CP_m$ in
     figure (\ref{det}) or e.g. in the anti-ferro-magnetic Ising model
     c.f.\cite{gross174}.
\item Finally, there is further a {\bf multi-critical point} $P_m$
  where more than two phases become indistinguishable. This is at the
  branching of several lines in the \{$e,n$\}-phase-diagram with
  $\lambda_1=0$,
  {\boldmath$\vecbm{\nabla}$}\mbox{$\lambda_1$}{\boldmath$= 0$}.
  Fig.~\ref{det} gives an illustration of a multi-critical point in a
  small system.
\end{itemize}

\section{Geometric foundation of irreversibility and the Second 
Law of Thermodynamics}

In the next three sections I want to deduce irreversibility and the
Second Law of statistical mechanics from the fundamental, microscopic,
reversible Newton mechanics of the $N$-interacting-particle dynamics.
I apologize this requires some more (very little) mathematics.
However, I believe this is still much simpler than any alternative
derivation proposed so far.

After succeeding to deduce equilibrium statistics including all
phenomena of phase transitions from Boltzmann's principle alone, even
for ``Small'' systems, i.e.\ non-extensive many-body
systems~\cite{gross174}, it is challenging to explore how far this
``most conservative and restrictive way to
thermodynamics''~\cite{bricmont00} is able to describe also the {\em
approach} of (possibly ``Small'') systems to equilibrium and the
Second Law of Thermodynamics. Thermodynamics describes the
development of {\em macroscopic} features of many-body systems
without specifying them microscopically in all details.

Before going into details I want to state the Second Law of
Thermodynamics in its perhaps most transparent way as follows, c.f.
Gallavotti~\cite{gallavotti99}, page32: Entropy as defined by
Boltzmann's principle (eq.\ref{boltzmentr1}) of an {\em isolated
many-body system approaching equilibrium either increases or remains
constant.}

One of the most piercing arguments against Boltzmann's statistical
explanation of the Second Law is due to
Zermelo~\cite{zermelo96,zermelo97}: A Hamiltonian system moves on a
closed trajectory in its $N$-body phase-space and returns after a
long recurrence time, the Poincar\'{e} recurrence time. So after some
time entropy should decrease again. Boltzmann's answer was: As this
time is astronomically large for a usual macroscopic systems ($\sim
10^{23}$ particles) these recurrences are irrelevant for practical
life. However, for a small system with a few tens of particles this
time becomes relevant and a reinvestigation of Zermelo's objection is
necessary.  Moreover, this question is also of fundamental
importance: Is irreversibility and the Second Law only due to the
extremely long recurrence times (Boltzmann) of macroscopic systems or
are these a general property of the basic {\em probabilistic} nature
of thermo-statistics?  Existing proofs of irreversibility out of the
microscopic time-reversible dynamics are using infinitely sized
systems and thus put Zermelo's objection aside.

\subsection{Measuring a macroscopic observable from a microscopic 
point of view}
\label{EPSformulation}
Before I address the Second Law, I have to clarify what I mean with
the label ``macroscopic observable''. A single point
$\{q_i(t),p_i(t)\}_{i=1,\cdots,N}$~\footnote{The curly brackets
  indicate the whole set of $6N$ coordinates $q_i,p_i$ of all
  particles $i$} in the $N$-body phase space corresponds to a detailed
specification of the system with all degrees of freedom ($dof$'s)
completely fixed at time $t$, i.e. a microscopic determination.
Fixing only the total energy $E$ of an $N$-body system leaves the
other ($6N-1$)-degrees of freedom unspecified.  A second system with
the same energy is most likely not in the same microscopic state as
the first, it will be at another point in phase space, the other
$dof$'s will be different.  I.e.\ the measurement of the total energy
$\hat{H}_N$, or any other macroscopic observable $\hat{M}$,
determines a ($6N-1$)-dimensional {\em sub-manifold} ${\cal{E}}$ or
${\cal{M}}$ in phase space. (The manifold ${\cal{M}}$ is called by
Lebowitz a {\em macro-state}~\cite{lebowitz99,lebowitz99a} which
contains $\Gamma_M=W(M)$ micro-states. I, however, prefer to use the
name ``state'' only for micro-states or points in phase space.)  All
points (the micro-states) in $N$-body phase space consistent with the
given value of $E$ and volume $V$, i.e.\ all points in the
($6N-1$)-dimensional sub-manifold ${\cal{E}}(N,V)$ of phase space are
equally consistent with this measurement.  ${\cal{E}}(N,V)$ is the
micro-canonical ensemble. This example tells us that {\em any
macroscopic measurement is incomplete and defines a sub-manifold of
points in phase space not a single point}. An additional measurement
of another macroscopic quantity $\hat{B}\{q,p\}$ reduces ${\cal{E}}$
further to the cross-section ${\cal{E}\cap\cal{B}}$, a
($6N-2$)-dimensional subset of points in ${\cal{E}}$ with the volume:
\begin{equation}
W(B,E,N,V)=\frac{1}{N!}\int{\left(\frac{d^3q\;d^3p}
{(2\pi\hbar)^3}\right)^N\epsilon_0\delta(E-\hat H_N\{q,p\})\;
\delta(B-\hat B\{q,p\})},
\label{integrM}\end{equation}

If $\hat H_N\{q,p\}$ as well as also $\hat B\{q,p\}$ are continuous
differentiable functions of their arguments, which I assume in the
following, then ${\cal{E}}\cap{\cal{B}}$ is closed. In the following I
use $W$ for the Riemann or Liouville volume (Hausdorff measure) of a
many-fold.

Micro-canonical thermo{\em statics} gives the conditional probability
$P(B|E,N,V)$ to find the $N$-body system in the sub-manifold
${\cal{E}}(N,V)\cap{\cal{B}}(N,V)$:
\begin{equation}
P(B|E,N,V)=\frac{W(B,E,N,V)}{W(E,N,V)}=e^{\ln[W(B,E,N,V)]-S(E,N,V)}
\label{EPS}\end{equation}
This is what Krylov seems to have had in mind~\cite{krylov79} and what
I will call the ``ensemble probabilistic formulation of Statistical
Mechanics ($EPS$) ''~\cite{gross183}. 

Similarly thermo{\em dynamics} describes the development in time of
some macroscopic observable $\hat{B}\{q_t,p_t\}$ of systems which
were specified at an earlier time $t_0$ by another macroscopic
measurement $\hat{A}\{q_0,p_0\}$.  It is related to the volume of the
sub-manifold
${\cal{M}}(t,t_0)={\cal{A}}(t_0)\cap{\cal{B}}(t)\cap{\cal{E}}$:
\begin{equation} W(A,B,E,t)=\frac{1}{N!}\int{\left(\frac{d^3q_t\;d^3p_t}
{(2\pi\hbar)^3}\right)^N\delta(B-\hat B\{q_t,p_t\})\;
\delta(A-\hat A\{q_0,p_0\})\;\epsilon_0\;\delta(E-\hat H\{q_t,p_t\})},
\label{wab}
\end{equation}
where $\{q_t\{q_0,p_0\},p_t\{q_0,p_0\}\}$ is the set of trajectories
solving the Hamilton-Jacobi equations
\begin{equation}
\dot{q}_i=\frac{\partial\hat H}{\partial p_i},\hspace{1cm}
\dot{p}_i=-\frac{\partial\hat H}{\partial q_i},\hspace{1cm}i=1\cdots N
\end{equation}
with the initial conditions $\{q(t=t_0)=q_0;\;p(t=t_0)=p_0\}$.  

For a large system with $N\sim 10^{23}$ the probability to find a
given value $B(t)$, $P[B(t)]$, is usually sharply peaked as function
of $B$ at its typical value. Such systems are called self-averaging.
Ordinary thermodynamics treats systems in the thermodynamic limit
$N\to\infty$ and gives only $<\!\!B(t)\!\!>$. However, here we are
interested to formulate the Second Law for ``Small'' systems i.e.\ we
are interested in the whole distribution $P[B(t)]$ not only in its
mean value $<\!\!B(t)\!\!>$. There are also many situation where the
system is not self-averaging, where a finite variance remains even in
the thermodynamic limit. (E.g.\ at phase transitions of first order
the energy {\em per particle} fluctuates in the canonical ensemble by
the specific latent heat.)

There is an important property of macroscopic measurements: Whereas at
finite times Hamilton dynamics evolves a compact region of phase space
again into a compact region, this does not need to be so at infinite
times.  Then, at $t\to\infty$, the set may not be closed anymore
(perhaps a fractal, see below). This means there exist series of
points $\{a_n\}\in{\cal{A}}(t=\infty)$ which converge to a point
$\lim_{n\to\infty}a_n=:a_{n=\infty}$ which is {\em not} in
${\cal{A}}(t=\infty)$. E.g.\ such points
$a_{n=\infty}\notin{\cal{A}}(\infty)$ may have intruded from the phase
space complementary to ${\cal{A}}(t_0)$.  Illustrative examples for
this evolution of an initially compact sub-manifold into a fractal set
are the generalized baker transformations discussed in this context by
ref.~\cite{fox98,gilbert00}. See reference~\cite{crc99} for the
fractal distribution produced by the general baker transformation. (As
any housewife knows, a baker dough becomes an infinitely thin
(fractal) puff pastry after pounding and folding it infinitely often.)
Only with {\em infinite resolution} this fractal distribution in phase
space can be seen. No macroscopic (incomplete) measurement can resolve
$a_{n=\infty}\notin{\cal{A}}(t=\infty)$ from its immediate neighbors
$a_n\in{\cal{A}}(t=\infty)$ in phase space with distances
$|a_n-a_{n=\infty}|$ less then any arbitrary small $\delta$.  In other
words, {\em at the time $t\to\infty$ no macroscopic measurement with
  its incomplete information about $\{q_{t=\infty},p_{t=\infty}\}$ can
  decide whether
  $\{q_0\{q_{t=\infty},p_{t=\infty}\},p_0\{q_{t=\infty},
  p_{t=\infty}\}\}\in{\cal{A}}(t_0)$ or not.} I.e.\ any macroscopic
theory like thermodynamics can only deal with the {\em closure} of
${\cal{A}}({t\to\infty})$. (The closure of a set of points ${\cal{A}}$
is defined as the set plus its limiting points $a_{n=\infty}$, also
called boundary points~\cite{crc99}).  If necessary, the sub-manifold
${\cal{A}}({t\to\infty})$ must be artificially closed~\footnote{First
  $t\to\infty$ then the closure, not the other way round c.f.\ 
  however, the discussion in the conclusion~\ref{conclus}.} to
$\overline{{\cal{A}}({t=\infty})}$ as developed further in
section~\ref{fractSL}.  {\em Clearly, in this approach this is the
  origin of irreversibility.}
\\~\\
Before going on, we must make a remark about what means infinite times
in reality. This is certainly a typical mathematical idealization:
There are several ``coarse graining'' processes due to which the
resolution of points in phase-space as demanded above is strongly
reduced: First due to quantum mechanics there is an ultimate coarse
graining over sizes of $\delta p \times \delta x \sim 2\pi\hbar$.
Second, and more important is the strong smearing discussed above due
to the highly reduced macroscopic information about the manifold
${\cal{A}}(t)$.  This of course, demands some detailed estimate
depending on the actual system.
\subsection{Fractal distributions in phase space, Second
Law}\label{fractSL} 

Let us examine the following Gedanken experiment: Suppose the
probability to find our system at points $\{q_t,p_t\}_1^N$ in phase
space is uniformly distributed for times $t<t_0$ over the sub-manifold
${\cal{E}}(N,V_1)$ of the $N$-body phase space at energy $E$ and
spatial volume $V_1$. At time $t>t_0$ we allow the system to spread
over the larger volume $V_2>V_1$ without changing its energy.  If the
system is {\em dynamically mixing}, the majority of trajectories
$\{q_t,p_t\}_1^N$ in phase space starting from points
$\{q_0,p_0\}_1^N$ with $q_0\in V_1$ at $t_0$ will now spread over the
larger volume $V_2$. As already argued by Gibbs~\cite{gibbs02,gibbs36}
the distribution ${\cal{M}}(t,t_0)$ will be filamented like ink in
water and will approach any point of ${\cal{E}}(N,V_2)$ arbitrarily
close.  $\lim_{t\to\infty}{\cal{M}}(t,t_0)$ becomes dense in the new,
larger ${\cal{E}}(N,V_2)$. (That is what ``mixing''
means~\cite{sinai70}.) The {\em closure}
$\overline{{\cal{M}}(t=\infty,t_0)}$ becomes equal to
${\cal{E}}(N,V_2)$.  This is clearly expressed by
Lebowitz~\cite{lebowitz99a,lebowitz99}.  Of course the Liouvillean
measure of the distribution ${\cal{M}}(t,t_0)$ in phase space at
$t>t_0$ will remain the same
($=tr[{\cal{E}}(N,V_1)]$)~\cite{goldstein59}:
\begin{eqnarray}
\left.tr[{\cal{M}}(t,t_0)]
\right|_{\{q_0\in V_1\}}
&=&\int_{\{q_0\{q_t,p_t\}\in
V_1\}}{\frac{1}{N!}\left(\frac{d^3q_t\;d^3p_t}
{(2\pi\hbar)^3}\right)^N\epsilon_0\delta(E-\hat
H_N\{q_t,p_t\})}\nonumber\\ &=&\int_{\{q_0\in
V_1\}}{\frac{1}{N!}\left(\frac{d^3q_0\;
d^3p_0}{(2\pi\hbar)^3}\right)^N\epsilon_0\delta(E-\hat
H_N\{q_0,p_0\})},\\
\mbox{because of: }\frac{\partial\{q_t,p_t\}}{\partial\{q_0,p_0\}}&=&1,
\label{liouville}\end{eqnarray}
(The label $\{q_0\in V_1\}$ of the integral means that the positions
$\{q_0\}_1^N$ are restricted to the volume $V_1$, whereas the momenta
$\{p_0\}_1^N$ are unrestricted.)

\begin{figure}[h]
\begin{minipage}[h]{6cm}
\begin{center}$V_a$\hspace{2cm}$V_b$\end{center}
\vspace*{0.2cm}
\includegraphics*[bb = 0 0 404 404, angle=-0, width=5.7cm,
clip=true]{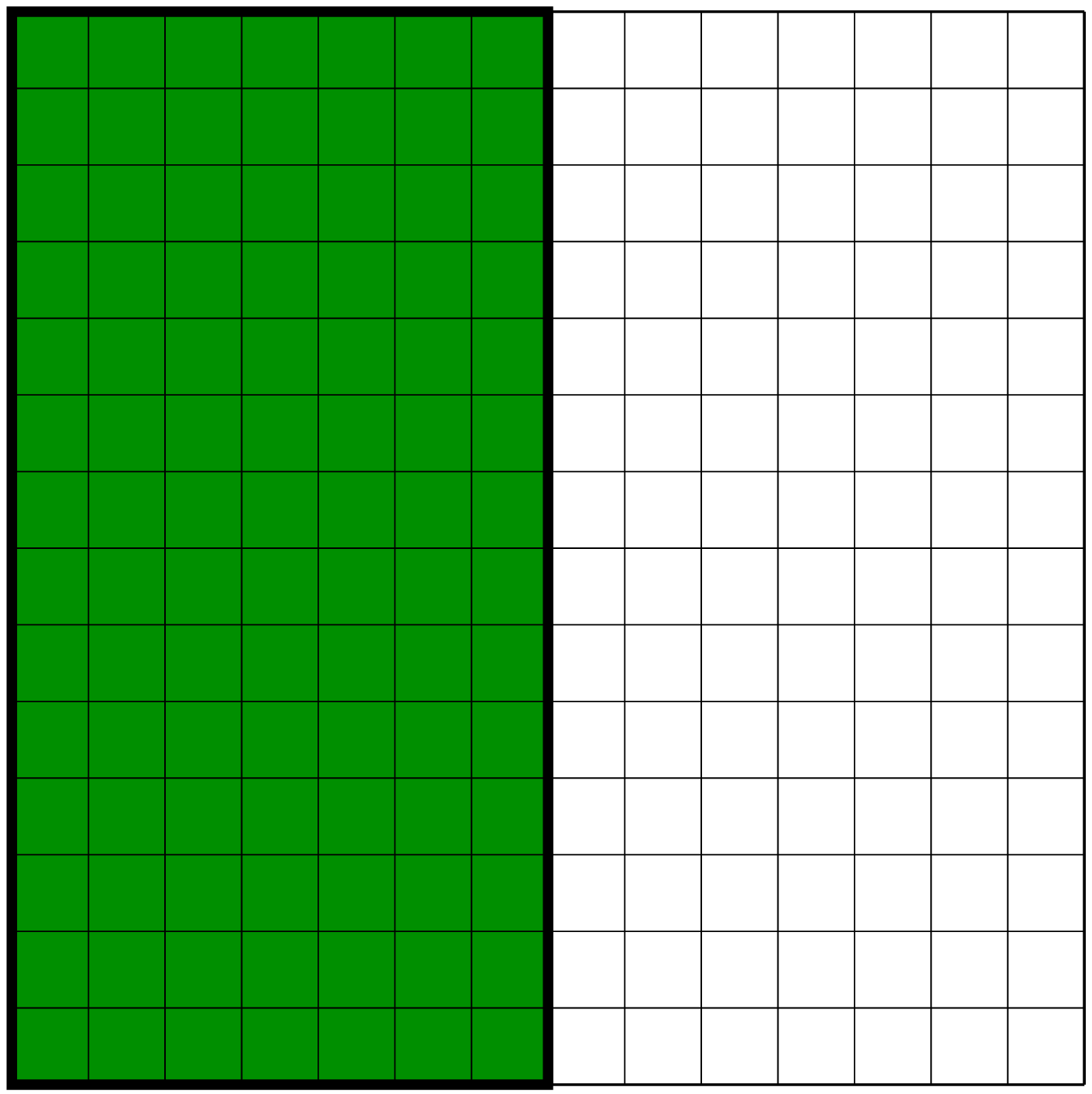}\begin{center}$t<t_0$\end{center}
\end{minipage}\lora\begin{minipage}[h]{6cm}
\begin{center}$V_a+V_b$\end{center}\vspace{-0.8cm}
\includegraphics*[bb = 0 0 490 481, angle=-0, width=6.9cm,
clip=true]{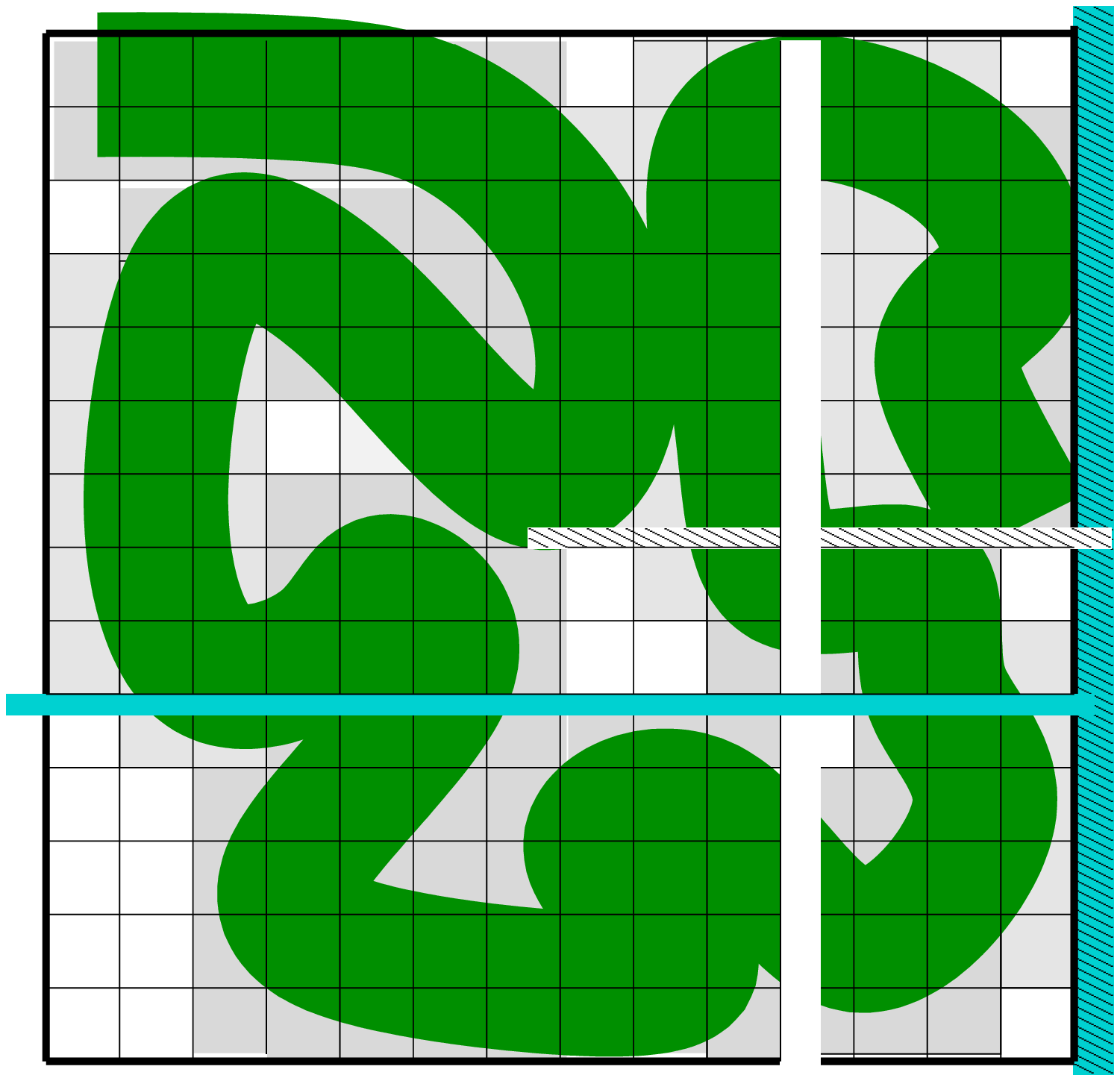}\begin{center}$t>t_0$\end{center}
\end{minipage}\\
\caption{The compact set ${\cal{M}}(t_0)$, left side, develops into an
  increasingly folded ``spaghetti''-like distribution
${\cal{M}}(t,t_0)$ in the phase-space with rising time $t$.  The
right figure shows only the early form of the distribution. At much
later times it will become more and more fractal and finally dense in
the new phase space.  The grid illustrates the boxes of the
box-counting method.  All boxes which overlap with ${\cal{M}}(t,t_0)$
contribute to the box-counting volume $\mbox{vol}_{box,\delta}$ and
are shaded gray.  Their number is $N_\delta$ \label{spaghetti}}
\end{figure}

In order to express this fact mathematically, I transform integrals
over the phase space like (\ref{integrM}) by changing to a new set
of orthogonal variables:
\begin{equation}
W(E,N,t,t_0)=
\frac{1}{N!}\int_{\{q_0\{q_t,p_t\}\subset V_1\}}{\left(\frac{d^3q_t\;d^3p_t}
{(2\pi\hbar)^3}\right)^N\epsilon_0\delta(E-\hat H_N\{q_t,p_t\})}
\end{equation}
into:
\begin{eqnarray}
\int{\left(d^3q_t\;d^3p_t\right)^N\cdots}&=&
\int{d\sigma_1\cdots d\sigma_{6N}\cdots}\\
d\sigma_{6N}&:=&\frac{1}{||\nabla\hat H||}
\sum_i{\left(\frac{\partial\hat H}{\partial
q_i}dq_i+\frac{\partial\hat H}{\partial p_i}dp_i\right)}=
 \frac{1}{||\nabla\hat H||}dE\\
||\nabla\hat H||&=&\sqrt{\sum_i{\left(\frac{\partial\hat H}{\partial
q_i}\right)^2+\sum_i{\left(\frac{\partial\hat H}{\partial p_i}\right)^2}}}\\
W(E,N,t,t_0)&=&\frac{1}{N!(2\pi\hbar)^{3N}}
\int_{\{q_0\{q_t,p_t\}\subset V_1\}}
{d\sigma_1\cdots d\sigma_{6N-1}
\frac{\epsilon_0}{||\nabla\hat H||}}.
\end{eqnarray}
Now, I {\em redefine} Boltzmann's definition of entropy
eq.(\ref{boltzmentr1} to \ref{phasespintegr}): by replacing the
Riemannian integral for $W$ by its {\em box-counting} ``measure'':
\begin{equation}
W(E,N,V)\to \displaystyle{B_d\hspace{-0.5 cm}
\int}_{\{q_0\{q_t,p_t\}\in V_1\}}
{d\sigma_1\cdots d\sigma_{6N-1}
\frac{\epsilon_0}{N!(2\pi\hbar)^{3N}||\nabla\hat H||}}
\label{boxM},
\end{equation}
i.e. the volume of ${\cal{M}}$ by that of its closure
$\overline{\cal{M}}$.  In detail we perform the following steps:
\begin{equation}
M_\delta(t,t_0):=<\!G\!>_\delta
*\mbox{vol}_{box,\delta}[{\cal{M}}(t,t_0)],\label{boxM1}
\end{equation}
to obtain $\mbox{vol}_{box,\delta}[{\cal{M}}(t,t_0)]$ we cover the
$d$-dim.\ sub-manifold ${\cal{M}}(t,t_0)$, here with $d=(6N-1)$, of
the phase space by a grid with spacing $\delta$ and count the number
$N_\delta\propto \delta^{-d}$ of boxes of size $\delta^{6N}$, which
contain points of ${\cal {M}}(t,t_0)$. [The following example may
explain this: To measure the area $A$ of a sheet of paper in 3d-space
one may cover the 3d-space by a grid of 3d-boxes of size $\delta^3$.
Only about $N_\delta=A\delta^{-2}$ boxes cut the paper. The area of
the paper is then $A=\lim_{\delta\to 0}\delta^2\times N_\delta$.]  This
is illustrated by fig.(\ref{spaghetti}). Then
$\mbox{vol}_{box,\delta}[{\cal{M}}(t,t_0)]:= \delta^d
N_\delta[{\cal{M}}(t,t_0)]$ and $<\!G\!>_\delta$ is the average of
$\frac{\epsilon_0}{N!(2\pi\hbar)^{3N}||\nabla\hat H||}$ over these
non-empty boxes of size $\delta$. The \underbar{$\lim$}$_{\delta\to
  0}\mbox{vol}_{box,\delta}[{\cal{M}}(t,t_0)]$ is the box-counting
volume of ${\cal{M}}(t,t_0)$ which is the same as the volume of its
closure $\overline{{\cal{M}}(t,t_0)}$, see below:
\begin{eqnarray}
\mbox{vol}_{box}[{\cal{M}}(t,t_0)]&:=&
\underbar{$\lim$}_{\delta\to 0}
\delta^d N_\delta[{\cal{M}}(t,t_0)]\label{boxvol}\\
\mbox{with }\underbar{$\lim *$}&=&\inf[\lim *]\mbox{ and write
symbolically:} \nonumber\\
\lim_{\delta\to 0}M_\delta(t,t_0)&=&
\lim_{\delta\to 0}<\!G\!>_\delta
*\mbox{vol}_{box,\delta}[{\cal{M}}(t,t_0)]\nonumber
\\&=:&
\displaystyle{B_d\hspace{-0.5 cm}
\int}_{\{q_0\{q_t,p_t\}\in V_1\}}
{\frac{1}{N!}\left(\frac{d^3q_t\;
d^3p_t}{(2\pi\hbar)^3}\right)^N\epsilon_0\delta(E-\hat
H_N)},
\end{eqnarray}
where $\displaystyle{B_d\hspace{-0.5 cm}\int}$ means that this
integral should be evaluated via the box-counting volume (the limit of
expression (\ref{boxM1}) with the use of (\ref{boxvol}) here with
$d=6N-1$. This is illustrated by Fig.\~(\ref{spaghetti}).

The volume of phase space covered by $M_\delta(t,t_0)$ is $\ge
W(E,N,V_1)$. Because the manifold remains compact for finite times and
because of Liouville's theorem eq.(\ref{liouville}), see also section
(\ref{EPSformulation}), we have
\begin{equation}
\underbar{$\lim$}_{\delta\to 0}M_\delta(t,t_0)=W(E,N,t_0,t_0)=W(E,N,V_1)
\end{equation}
At $t\to\infty$ the two limits $\delta\to 0,t\to\infty$ do in general
not commute and as assumed by Gibbs, in the case of a mixing
dynamics, the manifold ${\cal{M}}(t\to\infty)$ becomes dense in the
new micro-canonical manifold ${\cal{E}}(V_2)$.  Then
\begin{equation}
\underbar{$\lim$}_{\delta\to 0}\lim_{t\to\infty}M_\delta(t,t_0)=W(E,N,V_2)
> W(E,N,V_1).
\end{equation}
{\em This is the Second Law of Thermodynamics.} 

The box-counting is also used in the definition of the Kolmogorov
entropy, the average rate of entropy gain~\cite{crc99,falconer90}. The
box-counting ``measure'' is analogous to the standard method to
determine the fractal dimension of a set of points~\cite{falconer90}
by the box-counting {\em dimension}:
\begin{equation}
\dim_{box}[{\cal{M}}(t,t_0)]:=\underbar{$\lim$}_{\delta\to 0}
\frac{\ln{N_\delta[{\cal{M}}(t,t_0)]}}{-\ln{\delta}}.
\end{equation}

Like the box-counting dimension, the box-counting ``measure'' has the
peculiarity that it is equal to the measure of the smallest {\em
  closed} covering set. E.g.: The box-counting volume of the set
${\bf Q}$ of rational numbers between $0$ and $1$ is
$\mbox{vol}_{box}\{{\bf Q}\}=1$, and thus equal to the measure of the
{\em real} numbers, c.f.  Falconer~\cite{falconer90} section 3.1.
This is the reason why the box-counting ``measure'' is not a measure
in its mathematical definition because then we should have
\begin{equation}
\mbox{vol}_{box}\left[\sum_{i\in\{\bf  Q\}}({\cal{M}}_i)\right]=
\sum_{i\in\{\bf  Q\}}\mbox{vol}_{box}[{\cal{M}}_i]=0,\label{rational}
\end{equation}
therefore the quotation marks for the box-counting ``measure'', c.f.\ 
appendix~\ref{app}. 

Coming back to the the end of section (\ref{EPSformulation}), the
volume $W(A,B,\cdots,t)$ of the relevant ensemble, the {\em closure}
$\overline{{\cal{M}}(t,t_0)}$ must be ``measured'' by something like
the box-counting ``measure'' (\ref{boxM}) with the box-counting
integral $\displaystyle{ B_d\hspace{-0.5 cm}\int}$, which must replace
the integral in eq.(\ref{phasespintegr}). Because the box-counting
volume is equal to the volume of the smallest {\em closed} covering
set, the new, extended, definition of the phase-space integral
eq.(\ref{boxM}) is for compact sets like the equilibrium distribution
${\cal{E}}$ identical to the old one eq.(\ref{phasespintegr}) and our
redefinition of the phase-space integral by box-counting changes
nothing for equilibrium statistics.  {\em Therefore, one can simply
replace the old (Riemannian) Boltzmann-definition of the number of
complexions (i.e.  micro-states) and with it of the entropy by the
new one (\ref{boxM}) of course with the understanding that the
closure operation should be done {\em after} the times were
specified.} In this context it is interesting to notice that
Boltzmann was originally thinking of the phase-space discretized into
small but finite cells (!), see Gallavotti, page
41~\cite{gallavotti99}.

\section{Conclusion}\label{conclus}
The great conceptual clarity of micro-canonical thermo-statistics
compared to the grand-canon\-ical one is again clearly demonstrated.
This is due to the strict derivation of thermo-statistics from the
basic principles of mechanics, an old dream of Boltzmann. Essential
for this goal is the avoidance of the thermodynamic limit. Only then
phase transitions as the most interesting phenomena of thermodynamics
reveal their underlying physics, the creation of inhomogeneities and
interfaces. A further benefit of doing so is that the extended theory
applies to a much greater world: the non-extensive or ``Small''
sytems.

In this paper I showed that Boltzmann's principle
eq.(\ref{boltzmentr1}) covers in a simple and straight way both of
Lebowitz's central issues of statistical
mechanics~\cite{lebowitz99a}: The {\em appearance of phase
transitions} and the {\em geometrical origin of the Second Law}.
Earlier formulations of these ideas can be found
in~\cite{gross178,gross180}. Lebowitz emphasizes the necessity of
self-averaging for thermodynamics which describes the {\em typical}
outcome of a macroscopic measurement.  This can only be expected for
large systems, in the thermodynamic limit. However, {\em there are
many situations where even large systems are not self-averaging.}
E.g. at phase transitions of first order. Moreover, a whole world of
non-extensive systems, like the ``Small'' systems, show broad, often
not single peaked, phase-space distributions. An {\em extension} of
statistical mechanics to cover also these is demanded.

Macroscopic measurements $\hat{M}$ determine only a very few of all
$6N$ $dof$'s.  Any macroscopic theory like thermodynamics deals with
the {\em area} $W$ of the corresponding closed sub-manifold
$\overline{\cal{E}}$ in the $6N$-dim. phase space not with single
points. {\em Thermodynamics describes the behavior of averages over
this manifold.} The explicit averaging over ensembles, or finite
sub-manifolds in phase space, becomes especially important for the
micro-canonical ensemble of a {\em finite} or any other
non-self-averaging system.  E.g. in scattering experiments on nuclei
or atomic clusters an average over millions of events is taken. Thus
the whole distribution in the accessible phase space is explored. In
numerical simulations of phase transitions in finite systems this is
done by Monte Carlo averaging over the distribution in phase space.

Because of this necessarily coarsen information, macroscopic
measurements, and with it also macroscopic theories are unable to
distinguish fractal sets ${\cal{M}}$ from their closures
$\overline{\cal{M}}$. Therefore, I make the conjecture: the proper
manifolds determined by a macroscopic theory like thermodynamics are
the closed $\overline{\cal{M}}$.  However, an initially closed subset
of points at time $t_0$ does not necessarily evolve again into a
closed subset at $t=\infty$ {\em and the closure operation must be
explicitely done after setting the times in order to obtain a
quantity that is relevant for a macroscopic theory and can be
compared to thermodynamics}. As the closure operation and the
$t\to\infty$ limit do not commute, the macroscopic dynamics becomes
irreversible.

Here is the origin of the misunderstanding by the famous reversibility
paradoxes which were invented by Loschmidt~\cite{loschmidt76} and
Zermelo~\cite{zermelo96,zermelo97} and which bothered Boltzmann so
much~\cite{cohen97,cohen00}.  These paradoxes address to trajectories
of {\em single points} in the $N$-body phase space which must return
after Poincar\'{e}'s recurrence time or which must run backwards if
all momenta are exactly reversed.  Therefore, Loschmidt and Zermelo
concluded that the entropy should decrease as well as it was
increasing before. The specification of a single point in $6N$-dim
phase-space and the reversion of all its $3N$ momentum components
demands of course a {\em microscopic exact} specification of all $6N$
degrees of freedom not a determination of a few macroscopic degrees of
freedom only. As becomes clear from what was said above: {\em No
entropy is defined for a single point. This applies also to the
derived thermodynamic quantities like temperature $T=[\partial
S/\partial E]^{-1}$ or pressure $P=\frac{\partial S/\partial
V}{\partial S/\partial E}$.} Thermodynamics is addressed to the whole
manifold, ensemble of systems, with the same macroscopic constraints.
The ensemble develops irreversibly even though the underlying
Newtonian dynamics of each phase-space point is fully reversible. It
is highly unlikely that {\em all} points in the ensemble
${\cal{M}}(t,t_0)$ have commensurable recurrence times so that they
can return {\em simultaneously} to their initial positions.  {\em
Once the manifold has spread over the larger phase space it will
never return.}

Also other misinterpretation of Statistical Mechanics are pointed
out: The existence of phase transitions and critical phenomena are
{\em not} linked to the thermodynamic limit. They exist clearly and
sharply in ``Small'', non-extensive systems as well. As is
demonstrated by figure (\ref{det}), the micro-canonical phase diagram
shows much more details of the relevant phenomena of various phase
transitions than was possible up to now in the conventional canonical
approach.

Boltzmann's principle describes the equilibrium and the approach
towards the equilibrium of extensive as well of non-extensive
Hamiltonian systems.  By our derivation of micro-canonical
Statistical Mechanics for finite, eventually ``Small'' systems,
various non-trivial limiting processes are avoided.  Neither does one
invoke the thermodynamic limit of a homogeneous system with
infinitely many particles nor does one rely on the ergodic hypothesis
of the equivalence of (very long) time averages and ensemble
averages. As Bricmont~\cite{bricmont00} remarked Boltzmann's
principle is the most conservative way to Thermodynamics but more
than that it is the most straight one also.  {\em The single
axiomatic assumption of Boltzmann's principle, which has a simple
geometric interpretation, leads to the full spectrum of equilibrium
thermodynamics including all kinds of phase transitions and including
the Second Law of Thermodynamics.}

I take the fact serious that Thermodynamics as well as any other {\em
macroscopic} theory handles ensembles or sub-manyfolds and {\em not}
single points in phase-space. Thus the use of ensemble averages is
justified directly by the very nature of macroscopic (incomplete)
measurements. Entropy $s(e,n)$ is the natural measure of the
geometric size of the ensemble. With the Boltzmann definition of
$s(e,n)$, Statistical Mechanics becomes a {\em geometric} theory. The
topology of its curvature indicates all phenomena of phase
transitions independently of whether the system is ``Small'' or
large.  Coarse-graining appears as natural consequence of the
ensemble-nature.  The box-counting method mirrors the averaging over
the overwhelming number of non-determined degrees of freedom. Of
course, a fully consistent theory must use this averaging explicitly.
Presumably, the rise of the entropy can then already be seen at
finite times when the fractality of the distribution in phase space
is not yet fully developed. Then one would not depend on the order of
the limits $\lim_{\delta\to 0}\lim_{t\to\infty}$ as it was assumed
here. The coarse-graining is no more a mathematical ad hoc assumption.
It is the necessary consequence of the averaging over the $6N-M$
uncontrolled degrees of freedom. Moreover the Second Law in the
EPS-formulation of Statistical Mechanics is not linked to the
thermodynamic limit as was thought up to now
~\cite{lebowitz99a,lebowitz99}.

In this paper I did not contribute anything to the problem of
describing irreversible thermodynamics of stationary dissipative
systems as it is discussed e.g.\ by Gilbert and
Dorfman~\cite{gilbert99,gilbert00}, Rondoni and
Cohen~\cite{rondoni00}.  As mentioned already, dissipation does not
exist in the microscopic dynamics. It is not clear to me how far the
inclusion of dissipation predefines the arrow of time already which
should have been deduced from the theory.  The main problem for me was
the derivation of irreversibility from fully time reversible
microscopic dynamics under maximally clear conditions, i.e. of a
micro-canonical closed, finite system.
Gaspard~\cite{gaspard97,gaspard98} considers systems obeying a
dynamics that preserves the phase-space volume, i.e satisfying
Liouville's theorem, but under non-equilibrium steady state
conditions. Similarly to the present approach he had to coarse grain
(width $\delta$) the accessible phase space.  In conformity to the
standard view of thermodynamics being based on the thermodynamic
limit~\cite{lebowitz99} he then proves the rise of the entropy {\em
  after the limits} (in that order): first $V\to \infty$, then
$\delta\to 0$.  However, in this limit also the Poincar\'{e}
recurrence time becomes infinite and Zermelo's piercing argument
becomes blunted.  So in this approach Gaspard cannot treat our problem
of the Second Law in a {\em finite closed} Hamiltonian system which
seems to me to be the heart of the reversibility
paradox.\section{Acknowledgment} Thanks to A.Ecker for mathematical
advises and to J.Barr\'{e}, J.E.Votyakov and V.Latora who had useful
suggestions to improve the text. I am grateful to S.Abe, M.Baranger,
E.G.D.Cohen, P.Gaspard, H.Jaqaman, R.Klages, A.K.  Rajagopal and
M.Pettini for many illuminating discussions.
\section{Appendix}\label{app}
In the mathematical theory of fractals~\cite{falconer90} one usually
uses the Hausdorff measure or the Hausdorff dimension of the
fractal~\cite{crc99}.  This, however, would be wrong in Statistical
Mechanics.  Here I want to point out the difference between the
box-counting ``measure'' and the proper Hausdorff measure of a
manifold of points in phase space. Without going into too much
mathematical detail I can make this clear again with the same example
as above eq.(\ref{rational}): The Hausdorff measure of the rational
numbers $\in[0,1]$ is $0$, whereas the Hausdorff measure of the real
numbers $\in[0,1]$ is $1$. Therefore, the Hausdorff measure of a set
is a proper measure.  The Hausdorff measure of the fractal
distribution in phase space ${\cal{M}}(t\to\infty,t_0)$ is the same
as that of ${\cal{M}}(t_0)$, $W(E,N,V_1)$. Measured by the Hausdorff
measure the phase space volume of the fractal distribution
${\cal{M}}(t\to\infty,t_0)$ is conserved and Liouville's theorem
applies.  This would demand that thermodynamics could distinguish
between any point inside the fractal from any point outside of it
independently how close it is.  This, however, is impossible for any
macroscopic theory or experiment which has only macroscopic
information where all unobserved degrees of freedom are averaged
over.  That is the deep reason why the box-counting ``measure'' must
be taken and is a further origin for irreversibility.  \clearpage

\end{document}